\documentclass[12pt,emtex]{article}
\usepackage{amsfonts}
\usepackage{amssymb}
\usepackage{amscd}
\usepackage{amsmath}

\newcommand{\BEQ}{\begin{equation}}
\newcommand{\EEQ}{\end{equation}}
\def\bea{\begin{eqnarray}}
\def\eea{\end{eqnarray}}
\def\nn{\nonumber}

\newtheorem{Th}{Theorem}

\newtheorem{Lem}{Lemma}
\newtheorem{Prop}{Proposition}
\newtheorem{Cor}{Corollary}
\newtheorem{Hyp}{Hypothesis}
\newtheorem{Conjecture}{Conjecture}

\newtheorem{Remark}{Remark}

\newtheorem{Result}{Result}

\def\bea{\begin{eqnarray}}
\def\eea{\end{eqnarray}}

\def\bes{\begin{equation*} \begin{split}}
\def\ees{\end{split} \end{equation*}}

\def\C{{\mathbb{ C}}}
\def\CC{{\mathbb{ C}}}

\def\Cn{ {\mathbb C}^{n}  }

\def\d{\displaystyle}
\def\h{{\mathsf h}}

\def\one#1{#1^{\raise5pt\hbox{$\scriptstyle\!\!\!\!1$}}\,{}}
\def\two#1{#1^{\raise5pt\hbox{$\scriptstyle\!\!\!\!2$}}\,{}}

\def\t{\tau}
\def\g{\mathfrak{gl}_n}

\def\p{\partial_u}

\begin{document}

\begin{titlepage}
\hfill ITEP-TH-34/04 \vskip 2.5cm

\centerline{\LARGE \bf
Universal G-oper and Gaudin eigenproblem
}
\vskip 1.0cm
\vskip 1.0cm
\centerline{A. Chervov \footnote{E-mail:
chervov@itep.ru}
 D. Talalaev \footnote{E-mail:
talalaev@itep.ru}
}

\centerline{\sf Institute for Theoretical and Experimental Physics
\footnote{ITEP, 25 B. Cheremushkinskaya, Moscow, 117259, Russia.}}

\vskip 2.0cm

\begin{abstract}
This paper is devoted to the eigenvalue problem for the quantum Gaudin
system. We prove the universal correspondence between eigenvalues of Gaudin
Hamiltonians and the so-called $G$-opers without monodromy in general $\g$
case modulo a hypothesys on the
analytic properties of the solution of a KZ-type equation.
 Firstly we explore the quantum analog of the characteristic polynomial
which is a differential operator in a variable $u$ with the coefficients in
$U(\g)^{\otimes N}$. We will call it "universal $G$-oper".
It is constructed by the formula
$"Det"(L(u)-\partial_u)$ where $L(u)$ is
the quantum Lax operator for the Gaudin model and $"Det"$ is appropriate definition of
the determinant.
The coefficients of this
differential operator are quantum Gaudin Hamiltonians obtained by one of the
authors (D.T. hep-th/0404153). We establish the correspondence between eigenvalues and
$G$-opers as follows: taking
 eigen-values of  the Gaudin's hamiltonians  on the joint eigen-vector
in the tensor product of finite-dimensional representation of $\g$ and
substituting them into the universal $G$-oper we obtain the scalar
differential operator (scalar $G$-oper) which
conjecturally does not have monodromy.
We strongly believe that our quantization of the Gaudin model coincides
with quantization obtained from the center of universal enveloping algebra
on the critical level and that our scalar $G$-oper coincides with the
$G$-oper obtained by the geometric Langlands correspondence,
hence it provides very simple and explicit map (Langlands correspondence)
from Hitchin D-modules to $G$-opers in the case of rational base
curves. It  seems to be easy to generalize the constructions to the
case of other semisimple Lie algebras and models like XYZ.
\end{abstract}

\vskip 1.0cm
\end{titlepage}

\tableofcontents


\section{Introduction}
\subsection{Gaudin system}

Gaudin model was introduced  in \cite{Gaudin} (see section 13.2.2)
as a limit of the famous XXX-Heisenberg
model, which describes interaction of spins in a one dimensional chain.
Gaudin introduced his model for arbitrary semisimple Lie algebra,
while original physical motivation was restricted only to the $\mathfrak{sl}(2)$
case. Recently spin chains with algebras other then $\mathfrak{sl}(2)$
has found unexpected applications in gauge theories
(see \cite{BBGK} for recent comprehensive survey).
Gaudin model also attracts much attention in mathematics due
to its relation to geometric Langlands correspondence \cite{Fr1,Fr2},
Knizhnik-Zamolodchikov  and isomonodromy deformation
theory \cite{MAO-AML}, Hitchin  system \cite{ER1}, \cite{NN};
geometry of polygons \cite{Milson,Falqui}.

Let us recall the Gaudin construction and some notations which will be used
in the main text.
Let us denote by $\Phi$ the following $n\times
n$-matrix with coefficients in $\g$: we put element $e_{ij}$ on
$ij$-th place of the matrix. We also consider the direct sum $\g\oplus
... \oplus \g$ and denote by $\Phi_i$ the matrix defined as above but
with the elements from the  $i$-th copy of $\g$ in $\g\oplus ...
\oplus \g$. Let us introduce the Lax operator for the Gaudin model:
\bea \label{i-Lax-Gaud}
L(u)=\sum_{i=1...N} \frac{\Phi_i}{u-z_i}
\eea
where $a_i\in \CC$.

\bea
\label{i-tr-l2}
Tr L^2(u) =
\sum_i \frac{Tr \Phi_i^2}{(u-z_i)^2}
+\sum_k \frac{1}{u-z_k}
( \sum_{j\ne k} \frac{2 Tr \Phi_k \Phi_j}{(z_k-z_j)})
\eea

\bea
\label{i-ham-gaud}
H_k= \sum_{j\ne k} \frac{2 Tr \Phi_k \Phi_j}{(z_k-z_j)}
\mbox{ - are called quadratic Gaudin hamiltonians }\nn
\eea

To define the classical Gaudin model one needs to describe the
classical phase space of the system. This is taken to be
${\cal O}_1\times...\times {\cal O}_N$, where
${\cal O}_i$ are coadjoint orbits with Kirillov's symplectic form.
Functions $Tr L^k(z)$ can be restricted from $\g^*$
to coadjoint orbits and they give rise to completely integrable
hamiltonian system - the Gaudin system.

In the quantum case one considers the same formulas for the Lax operator and
quadratic hamiltonians only changing the generators of the Poisson
algebra $S(\g)^{\otimes N}$ by the same generators of
the universal enveloping algebra $U(\g)^{\otimes N}$.
It is well-known that $H_k$ commute on classical and quantum levels.
On classical level one can see from r-matrix technique that
$Tr L^k(z)$ and $Tr L^p(w)$ Poisson commute for arbitrary $k,p,z,w$.
So taking $Tr L^k(u)$ for $k=1,...,n$ and taking coefficients
of  these expressions at $\frac{1}{(u-z_i)^m}$ one obtains
the set of algebraically independent hamiltonians.
This is not true in general on the quantum level, for example one obtains:
$[Tr (L^4(z)), Tr (L^2(u))]\ne 0$
\cite{CRT}.

Mathematically speaking Gaudin hamiltonians generate
a large commutative subalgebra in $U(\g)^{\otimes N}$
which can be easily completed to maximal commutative subalgebra.

To define the quantum Gaudin model it is also necessary to describe
the Hilbert space of the model. This is taken to be (in accordance
to Kirillov's orbit method ideology) the tensor product of
some representations of $\g^{\oplus N}$:
$V_{1}\otimes...\otimes V_{N}$. Hamiltonians $H_k$
act on this Hilbert space and the quantum problem consists in finding
their spectrum, matrix elements and so on.

The integrability on the quantum level
is a corollary of the result of  Feigin and Frenkel \cite{Feigin}.
They proved the existence of a large
center for universal enveloping algebra of the Kac-Moody algebra
on the critical level. However their result does not provide explicit
formulas for generators of the center on critical level
(due to problems with normal ordering),
nor the explicit expressions for quantum hamiltonians of the Gaudin model.

The explicit formula for quantum commuting hamiltonians for
the Gaudin model was found in \cite{T}:

\bea \label{qi1}
\boxed{QI_i(u)=Tr_{1,\ldots,n}A_n (L_1(u)-\partial_u)(L_2(u)-\partial_u)\ldots
(L_i(u)-\partial_u) \mathbf{1}}
\eea
where $L(u)$ is the Lax operator for the Gaudin model, i.e. it lies in
$Mat_{n\times n}\otimes U(\g)^{\otimes N}$, $A_n$ is the
antisymmetrizer in $(\C^n)^{\otimes n}$, the notation $L_i$ implies that
matrix components of $L(u)$ lie
in $i$-th component of the tensor product of $Mat_{n\times n}^{\otimes n}$.

This result was obtained using
the Bethe commutative subalgebra in Yangian algebra $Y(\g)$.
This subalgebra was invented in Faddeev's school,
it is mentioned in \cite{Skl95} formula 3.16
and under the name "Bethe subalgebra"  appeared in \cite{NO}
(where it was attributed to Kulish,Sklyanin,Kirillov,Reshetikhin).


{\Remark~} Let us mention that the $\g$-valued matrices
$\Phi$ has been also considered in \cite{Kir} and interesting
conjectures about them were formulated.

One can take the other phase space for the same hamiltonians:
the direct product of $T^{*}GL_n$, then the obtained model will
be the Hitchin model on the curve $CP^1$ with $N/2$ double-points,
mixing these two phase spaces together and using reduction
one can obtain the trigonometric Calogero model (see  \cite{NN,CT1}).

\subsection{Main results}

Let us introduce
differential operators with values in $U(\g)^{\otimes N}$
\bea \label{univ-g-oper}
QD_i(u)=
Tr_{1,\ldots,n}A_n (L_1(u)-\partial_u)(L_2(u)-\partial_u)\ldots
(L_i(u)-\partial_u)
\eea
which in the case $i=n$
$$QD_n(u)="Det"(L(u)-\partial_u)$$
we will call
the {\it universal $G$-oper} by the reasons explained below. Here the usual
notations of formula (\ref{qi1}) are used. One can easily see that $Tr A_n
X_1...X_n=Det(X)$ in the case of a matrix $X$ with commutative entries.
The expression above is a kind of "quantization" for the characteristic polynomial:
$Det(L(u)-\lambda)$. In the theory of integrable systems this can also be
called "quantum spectral curve". The fact that $\lambda$ has been
substituted by the $\partial_u$ is related to the fact that in separated
variables the Poisson bracket is $\{u,\lambda\}=1$.

{\Remark~} One can see that for $2\times 2$ matrix
$$X= \left( \begin{array}{cc} a&b \\ c&d \end{array} \right).$$
the deformed determinant is just the symmetrized ordinary determinant
$$Tr A_2 (X_1)(X_2)=
\frac{1}{2}((Tr X)-TrX^2)=\frac{1}{2}(ad+da-bc-cb)$$
(pay attention that in our case the symmetrization is more
complicated due to the dependence on
$u$ and $\partial_u$).

The first important theorem (see section \ref{sect-g-oper})
consists in the following:
{\Result  \label{QChar_intro} All coefficients of the differential operator $QD_n(u)$ are quantum
hamiltonians of the Gaudin system, and the following formula holds:
\bea
\label{qd_intro}
QD_n(u)&=&QI_n(u)-nQI_{n-1}(u)\p+\frac {n(n-1)} 2 QI_{n-2}(u)\p^2-\ldots\nn\\
&=&\sum_{i=0}^{n} (-1)^i C_n^i QI_{n-i}(u)\p^i.
\eea
}

The second result is:
{\Result  \label{Mon1-intro} Substitution of coefficients  $QD_n(u)$
by  their images in the tensor product of finite-dimensional representations
of $GL(n)$ gives matrix-valued differential operator with
singularities at points $u=z_i$, which however does
not have monodromy modulo a hypothesys formulated below
on the analytic properties of the
KZ-type equation.
}

As a corollary one obtains the following:
{\Result  \label{Mon2-intro} Substitution of coefficients  $QD_n(u)$
by their  eigen-values on their joint eigen-vector
gives a scalar differential operator (scalar $G$-oper)
without monodromy.
}
The construction and the conjecture above  were anticipated in the
literature. The origin of such claims goes back to R. Baxter, M. Gaudin, further
E. Sklyanin has related this to the separation of variables. Recently some
close results were obtained by E. Frenkel, E. Mukhin, V. Tarasov and A.
Varchenko \cite{FMTV}.
Our main source of knowledge and insparation on the subject
is \cite{Fr1} and the conjecture in this form is motivated by this paper.

The importance of this result is based on the following fact: in this way
one can easily obtain the Bethe ansatz equation i.e. to find the spectrum of
the Gaudin model. We will do it in the next paper. The idea is explained in
\cite{Fr1} where the scalar $G$-oper was described explicitly only  for
$\mathfrak{gl}_2$,
$\mathfrak{gl}_3$.

Motivated by the analogy with Frenkel's paper we strongly
believe that result \ref{Mon2-intro} is true
in both directions: the scalar $G$-oper does not have
monodromy if and only if its coefficients are eigen-values
for the Gaudin hamiltonians.

These results are strongly connected with the geometric
Langlands correspondence of Beilinson and Drinfeld,
and Sklyanin's separation of variables theory,
as it is explained in \cite{Fr1}. We recall these
connections in section \ref{sect-Lang}.

{\Remark~}
In the Yangian case (it is related to the XXX spin-chain)
we also introduce the concept of the universal $G$-oper:

\bea
UG^{Yang} = Tr A_n
(e^{\d-\h\p}T_1(u)-1)(e^{\d-\h\p}T_2(u)-1)\ldots(e^{\d-\h\p}T_n(u)-1)
\eea

This expression corresponds also to the "quantization" of the characteristic polynomial,
but here  $\lambda$ is
quantized to $e^{\d-\h\p}$. This is in full agreement with
the Poisson bracket between separated variables $u$ and $\lambda$.

One has an obvious known fact:
{\Result  \label{QChar-yang-intro} All coefficients of the difference
operator $UG^{Yang}(u)$ are quantum
commuting hamiltonians of the XXX spin chain system.
}

By analogy with the Gaudin case we state the following:

{\Conjecture  Scalar difference operator
obtained from  $UG^{Yang}(u)$
by substituting its coefficients by their
eigen-values on their joint eigen-vector in
finite-dimensional representation of Yangian
has only univalued solutions.
}

The scalar discrete $G$-opers were introduced recently
by Mukhin and Varchenko see refrence 2 in \cite{FMTV},
we believe that our scalar Yangian $G$-oper
coincides with their.

\subsection{The plan}

The proof of result \ref{QChar_intro} is based
on the ideas of \cite{T} and use of Yangian technique,
especially Bethe subalgebra.

The proof of monodromy properties of the $G$-oper
takes the following lines:
\begin{itemize}
  \item We prove the formula
\bea
A_n (L_1(u)-\partial_u)\ldots(L_i(u)-\partial_u) A_n
=
n! A_n (L_1(u)-\partial_u)\ldots(L_i(u)-\partial_u)
\eea
With the help of a solution of
\bea \label{eqn-s}
(\partial_u - L(u)) \tilde S(u)=0
\eea
we define
\bea
\Psi(u)= A_n(v_1\otimes...\otimes v_{N-1}\otimes \tilde S(u))
\eea
which provides a solution for
\bea
QD_n \Psi(u)=0
\eea
for $v_i$ - generic vectors in $\CC^n.$ Moreover the corresponding linear
differential equation has regular singularities and hence the monodromy is
controlled by residues of $L(u)$ considered as elements of
$Mat_{n\times n}\otimes U(\g)^{\otimes N}.$
  \item Unfortunetely our case is exactly the resonance case and we only
state a hypothesys that the linear equation has no monodromy which we have
verified in some particular cases.
\item The final observation consists in restricting the global solution of
the generalized $G$-oper to the eigen vector of Gaudin hamiltonians
obtaining the scalar $G$-oper without monodromy.
\end{itemize}

{\bf Acknowledgements.}
We are indebted to A. Gorsky, S. Kharchev, M. Olshanetsky
and L. Rybnikov  for the discussions and
to A. Varchenko for pointing us the gap in the preliminary version of the paper.
The  work of both authors has been partially supported by the RFBR grant
04-01-00702.

\section{Universal $G$-oper and quantum commuting hamiltonians}
\label{sect-g-oper}
Let us recall standard Yangian notations.
$Y(\mathfrak{\g})$ is generated by
elements $t_{ij}^{(k)}$ subject to specific relations on the generating function
$T(u,\h)$ with values in $Y(\mathfrak{gl}_n)\otimes End(\mathbb{C}^n)$
$$T(u,\h)=\sum_{i,j} E_{ij}\otimes t_{ij}(u,\h),\qquad
t_{ij}(u,\h)=\delta_{ij}+\sum_k t_{ij}^{(k)} u^{-k}\h^k,$$ where $E_{ij}$ are matrices
with $1$ on the $i,j$-th place. The relations on this generating function
involve the Yang $R$-matrix $$R(u,\h)=1-\frac {\h} u \sum_{i,j}E_{ij}\otimes
E_{ji}$$ and take the RTT form:
$$R(z-u,\h)T_1(z,\h)T_2(u,\h)=T_2(u,\h)T_1(z,\h)R(z-u,\h)$$ in $End(\mathbb{C}^n)^{\otimes
2}\otimes Y(\mathfrak{gl}_n)[z,u,\h],$
where $$T_1(z,\h)=\sum_{i,j}E_{ij}\otimes {1}\otimes t_{ij}(z,\h),\quad
T_2(u,\h)=\sum_{i,j} {1}\otimes E_{ij}\otimes t_{ij}(u,\h).$$
There is a known realization of $Y(\g)$ in $U(\g)$ i.e.
a homomorphism $\rho_1:Y(\mathfrak{gl}_n)\rightarrow U(\mathfrak{gl}_n)$
\begin{equation}\label{gl}
T(u,\h)=1+\frac {\h} u \sum_{i,j}E_{ij}\otimes e_{ij}=1+
\frac \Phi u ,
\end{equation}
where $e_{ij}$ are
generators of $\g.$
The following expression also provides a realization of the Yangian in
$U(\mathfrak{gl}_n)^{\otimes k}$ for a given $k$-tuple
of complex numbers $\alpha=(z_1,\ldots,z_N)$
$$T^{\alpha}(u,\h)=T^1(u-z_1,\h)T^2(u-z_2,\h)\dots T^k(u-z_N,\h),$$ where $T^l(u-z_l,\h)$ is
the realization given by (\ref{gl}) with $e_{ij}$ lying in the $l$-th component of
$U(\mathfrak{gl}_n)^{\otimes N}.$ Let $\rho_{\alpha}$ be the corresponding
homomorphism $\rho_\alpha:Y(\mathfrak{gl}_n)\rightarrow U(\mathfrak{gl}_n)^{\otimes N}$.
\\
The {\em Bethe subalgebra} in $Y(\g)$ is generated by coefficients of the
following expressions
$$\tau_i(u,\h)=Tr A_n T_1(u,\h)T_2(u-\h,\h)\dots T_i(u-\h(i-1),\h)
\quad i=1,\ldots n. $$
One has
$$[\tau_i(u,\h),\tau_j(v,\h)]=0$$
\\
In \cite{T} it was considered formal expressions
\begin{equation}\label{sk}
s_i=\sum_{j=0}^{i}(-1)^j C_i^j\t_{i-j}
\end{equation}
where $\tau_0=Tr A_n 1=n!$ and it was proved that
$$s_i(u,\h)=\h^i QI_i(u)+O(\h^{i+1})$$
where $QI_i$ are quantum Gaudin hamiltonians given by the formula
\bea
\boxed{QI_i(u)=Tr_{1,\ldots,n}A_n (L_1(u)-\partial_u)(L_2(u)-\partial_u)\ldots
(L_i(u)-\partial_u) \mathbf{1}}\nn
\eea

{\Th \label{QChar} All coefficients of the differential operator $QD_n(u)$ given by
(\ref{univ-g-oper}) are quantum
hamiltonians of the Gaudin system, and the following formula holds:
\bea
\label{qd}
QD_n(u)&=&QI_n(u)-nQI_{n-1}(u)\p+\frac {n(n-1)} 2 QI_{n-2}(u)\p^2-\ldots\nn\\
&=&\sum_{i=0}^{n} (-1)^i C_n^i QI_{n-i}(u)\p^i.
\eea
}
\\
{\bf Proof~} Let us recall that the expression $QD_i$ can be obtained as a
coefficient at $\h^i$ (which is in fact the lowest non trivial coefficient)
of the following expression
\bea
&& Tr A_n
(e^{\d-\h\p}T_1(u)-1)(e^{\d-\h\p}T_2(u)-1)\ldots(e^{\d-\h\p}T_i(u)-1)\nn\\
&=&\sum_{j=0}^i\tau_j(u-\h,\h)(-1)^{i-j}C_i^j e^{\d -j\h\p} \label{cr}
\eea
Then expanding the exponential series and collecting terms with $\p^k$ we
obtain
\bea
\label{aux1}
\sum_{j=0}^n\tau_j(u-\h,\h)(-1)^{n-j}C_n^j \frac {(-j)^k\h^k} {k!} \p^k.
\eea
Next we observe the following:
\bea
\frac {j^k}{j!}=\sum_{m=1}^k B_{k,m} \frac 1 {(j-m)!}\nn
\eea
where the coefficients $B_{k,m}$ solve the system of recurrent relations
\bea
B_{l,m}=mB_{l-1,m}+B_{l-1,m-1}\nn
\eea
and do not depend on $j.$
Hence
\bea
C_n^j\frac {(-j)^k}{k!}=(-1)^k\sum_{m=1}^k B_{k,m}C_{n-m}^{n-j}\frac
{n!}{(n-m)!k!}\nn
\eea
Then the formula (\ref{aux1}) can be rewritten as follows:
\bea
\sum_{m=1}^k B_{k,m} \frac {n!}{(n-m)!k!}\left(
\sum_{j=0}^n
\t_j(u-\h,\h)(-1)^{n-j+k}C_{n-m}^{n-j}\right)\h^k\p^k\nn
\eea
Further one needs to note that
\bea
&&\sum_{j=0}^n
\t_j(u-\h,\h)(-1)^{n-j}C_{n-m}^{n-j}\\
&=&
Tr A_n
e^{\d-\h\p}T_1(u)\ldots e^{\d-\h\p}T_m(u)(e^{\d-\h\p}T_{m+1}(u)-1)
\ldots (e^{\d-\h\p}T_{n}(u)-1)\mathbf{1}\nn
\eea
and has the following expansion on $\h$
\bea
\h^{n-m}QI_{n-m}+O(\h^{n-m+1})\nn
\eea
Then collecting all terms of (\ref{aux1}) of lowest degree in $\h$ namely at
$\h^n$ and using the fact that $B_{k,k}=1$ we obtain the theorem
$\blacksquare$\footnote{We apologize for this too technical proof of this result,
we are sure that there exists much more simple and conceptual demonstration.}


\section{Solution and monodromy of $G$-oper}

\subsection{Antisymmetrization}
We construct a global solution of (\ref{qd}) in this section recalling that
this differential operator is obtained be taking the trace of the
lowest coefficients in $\h$ of a more simple
expression which provides a factorization:
\bea\label{chi}
\chi(u)&=&A_n
(e^{\d-\h\p}T_1(u)-1)(e^{\d-\h\p}T_2(u)-1)\ldots(e^{\d-\h\p}T_n(u)-1)\nn\\
&=&\h^n A_n (L_1(u)-\partial_u)(L_2(u)-\partial_u)\ldots
(L_n(u)-\partial_u)+O(\h^{n+1})
\eea

{\Prop
\bea\label{chi-2}
A_n
(e^{\d-\h\p}T_1(u)-1)(e^{\d-\h\p}T_2(u)-1)\ldots(e^{\d-\h\p}T_n(u)-1)
A_n =
\nn\\
n! A_n
(e^{\d-\h\p}T_1(u)-1)(e^{\d-\h\p}T_2(u)-1)\ldots(e^{\d-\h\p}T_n(u)-1)
\eea
}
\\
The proof of the proposition follows immediately from
the  lemmas below.
As a corollary  we obtain the similar proposition
for the case of the Gaudin Hamiltonians considering the coefficient at
$\h ^n$ of expression above.

{\Cor \label{Cor-chi-Gaudin}
\bea\label{chi-Gaudin}
A_n
(L_1(u)-\partial_u)(L_2(u)-\partial_u)\ldots
(L_n(u)-\partial_u)
A_n =
\nn\\
n! A_n
(L_1(u)-\partial_u)(L_2(u)-\partial_u)\ldots
(L_n(u)-\partial_u)
\eea
}

 {\Lem If $X (...\otimes a \otimes b \otimes ...) = - X  (...\otimes b
\otimes a \otimes ...) $, then $ n ! X= X A_n$, where $ (...\otimes a
\otimes b \otimes ...)  $ is an arbitrary vector in ${{({ \mathbb
C}^{n})}^{\otimes n}}$, $X  \in {\cal A} \otimes Mat_{n\times n}^{\otimes
n}$, $ {\cal A}$ is an arbitrary associative algebra, $Mat$ is the algebra of
matrices and $A_n \in Mat^{\otimes n} $ is the antisymmetrizer $ {{({\mathbb
C}^{n})}^{\otimes n}} \to {{({ \mathbb C}^{n})}^{\otimes n}}$. }
\\
{\bf Proof~} Arbitrary permutation  can be obtained
as the product of the transpositions exchanging only the neighbors $i$ and $i+1$,
so the conditions of the lemma above imply
$$X (v_1\otimes ... \otimes v_n )= (-1)^{sgn(\sigma)}
X(v_{\sigma (1)} \otimes ... \otimes v_{\sigma (n)}). $$ Hence
$$X A_n (v_1\otimes ... \otimes v_n ) =
X (\sum_{\sigma \in S_n} (-1)^{sgn(\sigma)}
(v_{\sigma (1)} \otimes ... \otimes v_{\sigma (n)})
= n ! X (v_1\otimes ... \otimes v_n )$$
$\Box$

{\Lem
\bea
 A_n
(e^{\d-\h\p}T_1(u)-1)(e^{\d-\h\p}T_2(u)-1)\ldots(e^{\d-\h\p}T_n(u)-1)
(...\otimes a \otimes b \otimes ...) \nn\\
= - A_n
(e^{\d-\h\p}T_1(u)-1)(e^{\d-\h\p}T_2(u)-1)\ldots(e^{\d-\h\p}T_n(u)-1)
(...\otimes b \otimes a \otimes ...)
\eea
}
\\
{\bf Proof~}
Recall the commutation relation for the Yangian:
$$ (1-\frac{\h P}{u-v}) T_1(u) T_2(v)= T_2(v)  T_1(u)
(1-\frac{\h P}{u-v}).$$
Then putting $u-v=\h$
one obtains:
$$A_2 T_1(u-\h) T_2(u-2\h ) = T_2(u-2\h) T_1(u-\h) A_2$$
which can be rewritten as
$$A_2  e^{\d-\h\p} T_1(u) e^{\d-\h\p} T_2(u ) =
 e^{\d-\h\p} T_2(u-\h) e^{\d-\h\p} T_1(u+\h) A_2.$$
One has also to note that
$$ A_2 (T_1(u)+ T_2(u))= (T_1(u)+ T_2(u)) A_2$$
because
\bea
A_2 (T_1(u)+ T_2(u)) &=& (1-P)(T_1(u)+ T_2(u))=
(T_1(u)+ T_2(u)) - P(T_1(u)+ T_2(u)) \nn\\
&=&
(T_1(u)+ T_2(u)) - (T_2(u)+ T_1(u)) P =
(T_1(u)+ T_2(u))(1-P).
\nn
\eea
Hence we obtain
\bea \label{something}
&&A_2 ( (e^{\d-\h\p}T_1(u)-1) (e^{\d-\h\p}T_2(u)-1) ) \nn\\
&=&A_2 ( e^{\d-\h\p}T_1(u) e^{\d-\h\p}T_2(u) -
e^{\d-\h\p}(T_1(u)+T_2(u)) +1)\nn\\
&=&
( e^{\d-\h\p}T_2(u-\h) T_1(u) -
e^{\d-\h\p}(T_1(u)+T_2(u)) +1) A_2 \nn\\
&=&\mbox{\em (by definition)~} B(u,\h,\p) A_2
\eea
\\
Let us denote by  $A_{i,i+1} = Id - P_{i,i+1} $ - the
antisymmetrizer in $i$-th and $i+1$-th place and also $B_{i,i+1}$ the
operator $B$ acting on the $i$-th and $i+1$-th tensor components.
Due to the formula \ref{something} we
see that
$$A_{i,i+1} ( (e^{\d-\h\p}T_i(u)-1) (e^{\d-\h\p}T_{i+1} (u)-1) )
= B_{i,i+1} A_{i,i+1} .$$
Recalling that the operator $ A_{i,i+1} $ commutes with
operators not acting on the $i$,$i+1$-th tensor components
of ${{({ \mathbb C}^{n})}^{\otimes n}}$ we obtain
\bea
&& A_n
(e^{\d-\h\p}T_1(u)-1)\ldots(e^{\d-\h\p}T_n(u)-1)
(...\otimes a \otimes b \otimes ...) \nn\\
&=&
\frac{1}{2} A_n A_{i,i+1}
(e^{\d-\h\p}T_1(u)-1)\ldots(e^{\d-\h\p}T_n(u)-1)
(...\otimes a \otimes b \otimes ...) \nn\\
&=&
\frac{1}{2} A_n
(e^{\d-\h\p}T_1(u)-1) \ldots (e^{\d-\h\p}T_{i-1}(u)-1)
\nn\\
&\times& A_{i,i+1}
(e^{\d-\h\p}T_{i}(u)-1)(e^{\d-\h\p}T_{i+1}(u)-1)
\ldots(e^{\d-\h\p}T_n(u)-1)
(...\otimes a \otimes b \otimes ...) \nn\\
&=&
\frac{1}{2} A_n
(e^{\d-\h\p}T_1(u)-1) \ldots (e^{\d-\h\p}T_{i-1}(u)-1)
B_{i,i+1}
\ldots(e^{\d-\h\p}T_n(u)-1)
\nn\\
&\times& A_{i,i+1}
(...\otimes a \otimes b \otimes ...) \nn\\
&=&
- \frac{1}{2} A_n
(e^{\d-\h\p}T_1(u)-1) \ldots (e^{\d-\h\p}T_{i-1}(u)-1)
B_{i,i+1}
\ldots(e^{\d-\h\p}T_n(u)-1)
\nn\\
&\times& A_{i,i+1}
(...\otimes b \otimes a \otimes ...) \nn\\
&=&
- A_n
(e^{\d-\h\p}T_1(u)-1)(e^{\d-\h\p}T_2(u)-1)\ldots(e^{\d-\h\p}T_n(u)-1)
(...\otimes b \otimes a \otimes ...)
\eea
The lemma is proved $\Box$
\\
The proposition follows immediately $\Box$

We propose here another technical lemma which was used in the first version
of the proof and can be useful in similar problems.
{\Lem If $A_n X = Y A_n$ then $A_n X A_n = n! A_n X $, where
$X,Y \in {\cal A} \otimes Mat_{n\times n}^{\otimes n}$,
$ {\cal A}$ is an arbitrary
associative algebra, $Mat$ is the algebra of matrices and
$A_n \in Mat^{\otimes n} $ is the
antisymmetrizer $ {{({\mathbb C}^{n})}^{\otimes n}} \to
{{({ \mathbb C}^{n})}^{\otimes n}}$.
}
\\
{\bf Proof} Recall that $ A_n^2=A_n n!$, so
$A_n X A_n = Y A_n A_n= Y A_n n!= n! A_n X $. $\Box$


{\Th
\label{antisymm}
Given a solution $S(u)$ with values in $(\C^n)^{\otimes n}\otimes V$
(where $V$ is a representation of $U(\g)^{\otimes k}$) of
the equation
\bea
\label{eqgen}
A_n(L_1(u)-\partial_u)(L_2(u)-\partial_u)\ldots
(L_n(u)-\partial_u)S(u)=0
\eea
we take $\Psi(u)$ with values in $V$ defined by
\bea
\label{psi}
A_n S(u)=\Psi(u)A_n(e_1\otimes\ldots\otimes e_n).
\eea
$\Psi(u)$ provides a solution of the equation
\bea
QD_n(u)\Psi(u)=0\nn
\eea
}
\\
{\bf Proof~~} Using corollary \ref{Cor-chi-Gaudin}
we see that
\bea
0 &=& n!  A_n(L_1(u)-\partial_u)(L_2(u)-\partial_u)\ldots
(L_n(u)-\partial_u)S(u)\nn\\
&=& A_n(L_1(u)-\partial_u)(L_2(u)-\partial_u)\ldots
(L_n(u)-\partial_u) \frac {A_n}{n!} A_n S(u)\nn\\
&=&
QD_n(u)\Psi(u) A_n(e_1\otimes\ldots\otimes e_n)\nn
\eea
where we have used that ${A_n}(n!)^{-1}$ is a projector.
Hence $ QD_n(u)\Psi(u)=0~$ $\Box$

\subsection{Monodromy properties}
Let us firstly recall the fact that quantum Gaudin hamiltonians can be
restricted to the invariant part
$V=(V_{\lambda_1}\otimes...\otimes V_{\lambda_n})^{(GL(n)~invariant)}$
of the tensor product of representation with respect to
the diagonal action and this space is usually considered as a representation
for the quantum Gaudin model.

{\Hyp
\label{KZ}
The equation
$$(L(u)-\partial_u)\tilde{S}(u)=0$$ for the
$\Cn\otimes V$-valued function $\tilde{S}(u)$ has a fundamental solution
without monodromy.
}
Modulo this hypothesys one obtain the following
{\Th
\label{th-monod}
 In the case of the representation space
$V=(V_{\lambda_1}\otimes...\otimes V_{\lambda_n})^{(GL(n)~invariant)}$
all the solutions of $QD_n(u)\Psi(u)=0$ have trivial monodromy
at infinity and at the points $z=z_i.$
}

{ \Remark If we do not take the invariant part of
$V_{\lambda_1}\otimes...\otimes V_{\lambda_n}$ the monodromy
at infinity is not zero, but is easily controllable.
}
\\
{\bf Proof ~} One can construct a
solution to the equation
$$(L_1(u)-\partial_u)(L_2(u)-\partial_u)\ldots
(L_n(u)-\partial_u)S(u)=0$$
just taking the fundamental solution of the equation
$$(L(u)-\partial_u)\tilde S(u)=0$$
and considering $S(u)=v_1\otimes...v_{n-1}\otimes \tilde S(u)$
where $v_i$ are arbitrary vectors from $\Cn.$ Then one has to apply Theorem
\ref{antisymm} to obtain solutions $\square$

{\Remark  In the case of other representations like Verma modules the monodromy is
expected to be
easily controllable. In the case of $sl_n$ the monodromy becomes trivial after
the mapping $GL_n\to PGL_n$ (recall that $SL_n$ and $PGL_n$ are Langlands
dual groups, so it is not accidental).
}

\subsection{The main correspondence}
Modulo the hypothesys \ref{KZ} one obtains:

{\Th The common eigen-vectors of the Gaudin hamiltonians correspond to the
scalar differential operators ("G-opers")
$ \sum_i h_i(u) \partial_u^i$ which do not have monodromy,
where $h_i(u)$ are values of the quantum hamiltonians
$(-1)^iC^i_n QI_i(u)$ on the joint eigen-vector.
}
\\
{\bf Proof} Following theorem \ref{th-monod}
one just need to restrict the differential operator
$$QD_n(u)=\sum_i (-1)^iC^i_n QI_i(u)\partial_u^i $$
to a common eigen-vector of the quantum Gaudin hamiltonians
$QI_i(u)$
$\Box$


\section{Relations}
\label{sect-Lang}

In this section we recall several relations of our principal subject
with the geometric Langlands correspondence of Beilinson and Drinfeld,
with the separation of variables technique, and annonce some
generalizations of the obtained results.

\subsection{Langlands correspondence}

The geometric Langlands correspondence predicts a bijection
between Hitchin D-modules on the moduli space of vector
bundles on the one side and local systems on the curve on the other
side.
Some of local systems can be realized as connections
having a flag of invariant subbundles - such connections
can be defined as differential operators called $G$-opers.
On the automorphic  (Hitchin)
side the Hitchin D-module is just the D-module generated by the
differential operators $\hat H_i-\C_i$ where $\hat H_i$
are quantum Hitchin hamiltonians and $C_i$ are arbitrary constants
parameterizing Hitchin D-modules.
So there should appear a correspondence between
some constants $C_i$ and some differential operators ($G$-opers).

The Langlands correspondence can be generalized to the
case of vector bundles with parabolic structures,
in this case instead of the Hitchin model appears the Gaudin model.

{\Conjecture Our quantum commuting Gaudin hamiltonians generate
the same D-module as the hamiltonians obtained from
the center of $\widehat {\g}$ on the critical level (these objects
 are usually considered in the geometric Langlands
correspondence theory).
}

Let us remark that the advantage of our approach is the explicitness
of proposed formulas in contrast to
the approach based on the center of
$\widehat {\g}$ on the critical level, which
still cannot be described explicitly except
of generators of degree $2.$

{\Conjecture Our scalar $G$-oper obtained by substituting
the coefficients of universal $G$-oper by
some constants $C_i$ is the same as the
Langlands correspondent $G$-oper of the Hitchin D-module
generated by $\hat H_i-\C_i$.
}

\subsection{Separated variables (Baxter equation)}

The relation of the $G$-oper to the separation of variables
is as follows:
in separated variables the wave functions
of the quantum system is factorized to the
product of one particle functions $\Psi=\Psi^{sv}(u_1)...\Psi^{sv}(u_n)$.
It appears that the differential (or difference) equation for the
wave functions $\Psi^{sv}$ (called Baxter equation)
is given by the scalar $G$-oper,
this is proved in Frenkel's paper for $\mathfrak{sl}_2$ Gaudin case
and seems to be true in general.
We hope to show it in our next publications.


\subsection{Generalizations}

The results presented here can be generalized in the following
directions: firstly one can consider the Gaudin model for other
semisimple Lie algebras $\mathfrak{g}$ - to do this one need to generalize the
notion of the matrix $\Phi.$ This can be done
by the consideration the universal $r$-matrix
corresponding to these Lie algebras and corresponding analogues
of the Bethe subalgebra.

Another generalization is the case of XXZ and XYZ models.
In these cases we expect the following change in the
formula for the universal $G$-oper
$$
UG^{XXZ-naive}(v)=
Tr A_n(L_n(exp(v))-exp(\d\partial_v))...(L_1(exp(u))-exp(\d\partial_v)).
$$
This formula probably  needs  further corrections.
It is motivated by two examples - the Gaudin and XXX model,
where we have seen the classical characteristic polynomial
$det(L(u)-\lambda)$ to be  "quantized" to the Universal $G$-oper by the
rule that $\lambda$ is changed to some operator $\hat{\lambda}$
such that the commutation relation between $u$ and
$\hat{\lambda}$ is the quantization of the Poisson bracket
between $u$ and $\lambda$ in separated variables.
Besides it is known for XXZ model that in separated variables
the Poisson  bracket $\{u,\lambda \} =u\lambda$.
In the same vein one can try to generalize the
formula for the universal $G$-oper by for XYZ model.
We believe that our propositions about the monodromy
can be generalized to these examples also and
hence they eventually will give a new approach
to the Bethe ansatz equations in these models.

Another kind of generalization concerns the quantization problem for the
integrable systems with higher poles Lax operator,
the so-called polynomial matrix models
\cite{Beauv,AHH} and related Hitchin models
on singular curves \cite{CT2} or for integrable systems obtained in the
framework of \cite{Chern}.
The formula for the universal
$G$-oper is expected to be absolutely the same.
But the results about the monodromy
are more problematic, possibly
all the information will be encoded in the Stokes
matrices, instead of monodromy ones,
this deserves further investigations.


\begin{thebibliography}{50}

\bibitem[Gaudin]{Gaudin}
M. Gaudin, {\em La Fonction d' Onde de Bethe}, Masson,
Paris (1983).

 M. Gaudin, {\em Diagonalisation d' une classe
d' hamiltoniens de spin}, J. de Physique 37, 1087-1098 (1976).


\bibitem[Belitsky04]{BBGK}
A.V. Belitsky, V.M. Braun, A.S. Gorsky, G.P. Korchemsky
{\em Integrability in QCD and beyond},
hep-th/0407232


\bibitem[Frenkel95]{Fr1}
 Edward Frenkel, {\em  Affine Algebras, Langlands Duality and Bethe Ansatz},
q-alg/9506003


\bibitem[Frenkel04]{Fr2}
 Edward Frenkel, {\em  Gaudin model and opers},
math.QA/0407524

\bibitem[LO97]{MAO-AML}
A.M.Levin, M.A.Olshanetsky,
{\em Classical limit of the Knizhnik-Zamolodchikov-Bernard equations as
hierarchy of isomonodromic deformations. Free fields approach},
hep-th/9709207


\bibitem[Enriquez95]{ER1}
B. Enriquez, V. Rubtsov, {\em Hitchin systems, higher Gaudin
operators and $r$-matrices.} Math. Res. Lett. {\bf 3} (1996)
343-357; alg-geom/9503010.

\bibitem[Nekrasov95]{NN}
N. Nekrasov, {\em Holomorphic bundles and many-body systems.}
PUPT-1534, Comm. Math. Phys.,{\bf 180} (1996) 587-604;
hep-th/9503157.

\bibitem[KM96]{Milson}
 M. Kapovich, J. Millson,  {\em The symplectic geometry
of polygons in Euclidean space},J. Differ. Geom. {\bf{44}}, 479--513 (1996)

\bibitem[FM03]{Falqui}
Gregorio Falqui, Fabio Musso,
Gaudin Models and Bending Flows: a Geometrical Point of View,
nlin.SI/0306005

\bibitem[Sklyanin95]{Skl95} E.K. Sklyanin, {\em Separations of variables:
new trends.} Progr. Theor. Phys. Suppl. {\bf 118} (1995), 35--60.
solv-int/9504001

\bibitem[FF92]{Feigin}
Feigin, Boris; Frenkel, Edward
{\em Affine Kac-Moody algebras at the critical level and Gelfand-Dikii algebras}. (English)
Int. J. Mod. Phys. A 7, Suppl. 1A, 197-215 (1992).

\bibitem[CRT04]{CRT}
A. Chervov, L. Rybnikov, D. Talalaev {\em
Rational Lax operators and their quantization}, hep-th/0404106

\bibitem[Talalaev04]{T}
D. Talalaev {\em Quantization of the Gaudin system}, hep-th/0404153

\bibitem[NO95]{NO}
Maxim Nazarov, Grigori Olshanski
{\em Bethe Subalgebras in Twisted Yangians},
q-alg/9507003

\bibitem[CT03]{CT1}
A. Chervov, D. Talalaev
{\em Hitchin system on singular curve I},
Theoretical and Mathematical Physics, 140(2):1043-1072(2004),
hep-th/0303069


\bibitem[Kirillov00]{Kir}
A.A. Kirillov, {\em Introduction to family algebras}, Moscow Math. Journal, No 1, 2000

\bibitem[FMTV]{FMTV}

E. Mukhin, A. Varchenko,
{\em Critical points of master functions and flag varieties}, math.QA/0209017

E. Mukhin, A. Varchenko,
{\em Miura Opers and Critical Points of Master Functions}, math.QA/0211321


E. Mukhin, A. Varchenko,
{\em Solutions to the XXX type Bethe ansatz equations and flag},  math.QA/0312406

Edward Frenkel,
{\em Opers on the projective line, flag manifolds and Bethe Ansatz},
math.QA/0308269

Evgeny Mukhin, Alexander Varchenko {\em
Discrete Miura Opers and Solutions of the Bethe Ansatz Equations},
math.QA/0401137

 Edward Frenkel, {\em  Gaudin model and opers},
math.QA/0407524

\bibitem[Beauville90]{Beauv}
A. Beauville, {\em Jacobiennes des Courbes Spectrales
et Systemes Hamiltoniens Completement
Integrables.}  Acta Math. 164 No.3/4 (1990) 211-235


\bibitem[AHHP]{AHH}
Adams, M.R.; Harnad, J.; Previato, E.
Isospectral Hamiltonian flows in finite and infinite dimensions. I: Generalized Moser systems and moment maps into loop algebras. (English)
[J] Commun. Math. Phys. 117, No.3, 451-500 (1988).

 M.R. Adams, J. Harnad, J. Hurtubise, {\em Darboux
Coordinates and Liouville-Arnold Integration in Loop Algebras.}
Commun. Math. Phys. {\bf 155} (1993), 385-413.

 M.R. Adams, J. Harnad, J. Hurtubise, {\em Darboux
   coordinates on coadjoint orbits of Lie algebras}. Lett. Math. Phys. {\bf
40} (1997), 41--57.

\bibitem[CT03-2]{CT2}
A. Chervov, D. Talalaev
{\em Hitchin system on singular curve II};
hep-th/0309059


\bibitem[Chernyakov03]{Chern} Yu. Chernyakov, {\em
Integrable systems obtained by the gluing points from rational and elliptic
Gaudin systems},  hep-th/0311027


\end{thebibliography}
\end{document}